\documentclass[runningheads]{llncs}
\usepackage{graphicx}
\begin{document}
\title{Improving of Robotic Virtual Agent's errors that are accepted by reaction and human's preference}
\titlerunning{RVA errors are accepted by reaction and human's preference}
%
\author{Takahiro Tsumura\inst{1,2}\orcidID{0000-0002-3145-3120} \and
Seiji Yamada\inst{2,1}\orcidID{0000-0002-5907-7382}}
\authorrunning{T. Tsumura et al.}
%
\institute{Department of Informatics, The Graduate University for Advanced Studies, SOKENDAI, Tokyo, Japan \and
Digital Content and Media Sciences Research Division, National Institute of Informatics, Tokyo, Japan}
\maketitle              
\begin{abstract}
One way to improve the relationship between humans and anthropomorphic agents is to have humans empathize with the agents. 
In this study, we focused on a task between an agent and a human in which the agent makes a mistake.
To investigate significant factors for designing a robotic agent that can promote humans empathy, we experimentally examined the hypothesis that agent reaction and human’s preference affect human empathy and acceptance of the agent's mistakes. 
The experiment consisted of a four-condition, three-factor mixed design with agent reaction, selected agent's body color for human's preference, and pre- and post-task as factors. 
The results showed that agent reaction and human’s preference did not affect empathy toward the agent but did allow the agent to make mistakes. 
It was also shown that empathy for the agent decreased when the agent made a mistake on the task. 
The results of this study provide a way to control impressions of the robotic virtual agent's behaviors, which are increasingly used in society.

\keywords{human-agent interaction \and empathy agent \and human's preference}
\end{abstract}
\section{Introduction}
Humans use a variety of tools in their daily lives. They become attached to these tools and sometimes treat them like humans. 
The Media Equation claims that humans treat artifacts like humans~\cite{Reeves96}. 
It has been shown that humans have the same feelings toward artifacts as they do toward other humans. 
In fact, there are examples of people empathizing with artifacts in the same way that humans empathize with humans. 
Typical examples include cleaning robots, pet-type robots, characters in competitive video games, and anthropomorphic agents that provide services such as online shopping and help desks. 
However, when humans and agents work together, such as at a help desk or when cleaning, there can be problems when humans view these agents as tools. 
There are also certain types of humans who cannot accept agents~\cite{Nomura06,Nomura08,Nomura16}. 
Currently, such agents are already being used in human society and coexist with humans.

Agents used in society often perform tasks with humans. 
At times, an agent may get the task wrong. When an agent makes a mistake on a task, many humans lower their expectations and trust in the agent. 
However, we often develop agents so that they do not make mistakes, but rarely do we take an approach that preserves the human's impression of the agent when it actually makes a mistake. 
One way to do so is to have the human empathize with the agent. 
When agents are used as tools, they may not need empathy, but when they are used in place of humans, being empathized with by humans can help build a smooth relationship.

Humans and anthropomorphic agents already interact in a variety of tasks. 
For a human to develop a good relationship with an agent, empathy toward the agent is necessary. 
Empathy makes it easier for humans to take positive action toward an agent and to accept it~\cite{Tsumura22,Tsumura23-1,Tsumura23-2}. 
This can also be effective when an agent makes a mistake on a task. 

Although various factors have been studied that cause empathy, including verbal and nonverbal information, situations, and relationships, this study focuses on situations in which the robotic virtual agent (RVA) gets the task wrong and experimentally examines how the agent's reaction and the agent's human's preference affect empathy.
The empathy investigated in this study is human empathy toward the agent, and we investigated changes in impressions of the agent used in the experiment.

\section{Related work}
In the field of psychology, empathy has been the focus of much attention and research. 
Omdahl~\cite{Omdahl95} classified empathy into three main categories: (1) affective empathy, which is an emotional response to another person's emotional state, (2) cognitive empathy, which is a cognitive understanding of another person's emotional state, and (3) empathy that includes both of the above.
Preston and De Waal~\cite{Preston02} proposed that at the heart of empathic responses is a mechanism that allows the observer access to the subjective emotional state of the subject. 
The Perception-Action Model (PAM) was defined by them to unify the differences in empathy. 
They defined empathy as a total of three types: (a) sharing or being affected by the emotional states of others, (b) evaluating the reasons for emotional states, and (c) the ability to identify and incorporate the perspectives of others. 

Various questionnaires are used as measures of empathy, but we used the Interpersonal Reactivity Index (IRI). 
IRI, also used in the field of psychology, is used to investigate the characteristics of empathy~\cite{Davis80}. 
There is another questionnaire, the Empathy Quotient (EQ)~\cite{Cohen04}, but we did not use it in our study because we wanted to investigate which categories of empathy were affected after experiencing the task.

In the fields of human-agent interaction (HAI) and human-robot interaction (HRI), empathy between humans and agents or robots is studied. 
The following studies have been conducted in various areas of HRI. 
Leite et al.~\cite{Leite14} conducted a long-term study in elementary schools to present and evaluate an empathy model for a social robot that interacts with children over a long period of time. 
They measured children's perceptions of social presence, engagement, and social support. Mathur et al.~\cite{Mathur21} present a first approach to modeling user empathy elicited during interaction with a robot agent. 
They collected a new dataset from a novel interaction context in which participants listen to a robotic storyteller. 
Johanson et al.~\cite{Johanson23} examined whether the use of verbal empathic statements and head nodding by a robot during video-recorded interactions between a healthcare robot and a patient could improve participants' trust and satisfaction.

In addition, the following studies have been conducted in the field of HAI. 
Okanda et al.~\cite{Okanda19} focused on appearance and investigated Japanese adults' beliefs about friendship and morality toward robots. 
They examined whether the appearances of robots (i.e., humanoid, dog-like, oval-shaped) differed in relation to their animistic tendencies and empathy. 
Samrose et al.~\cite{Samrose20} designed a protocol to elicit user boredom to investigate whether empathic conversational agents can help reduce boredom. 
With the help of two conversational agents, an empathic agent and a non-empathic agent, in a Wizard-of-Oz setting, they attempted to reduce the user's boredom. 
Al Farisi et al.~\cite{Al-Farisi22} believe that in order for chatbots to have human-like cues, it is necessary to apply the concepts of human-computer interaction (HCI) to chatbots and compare the empathy of two chatbots, one with anthropomorphic design cues (ADC), and one without. 
Tsumura and Yamada~\cite{Tsumura22} focused on tasks between agents and humans, experimentally examining the hypothesis that task difficulty and task content promote human empathy.
We also considered the design of empathy factors from previous studies of anthropomorphic agents using empathy. 
Tsumura and Yamada~\cite{Tsumura23-1} focused on self-disclosure from agents to humans in order to enhance human empathy toward anthropomorphic agents, and they experimentally investigated the potential for self-disclosure by agents to promote human empathy. 
Tsumura and Yamada~\cite{Tsumura23-2} also focused on tasks in which humans and agents engage in a variety of interactions, and they investigated the properties of agents that have a significant impact on human empathy toward them. 

Paiva defined the relationship between human beings and empathic agents, referred to as empathy agents, as designed in previous HAI and HRI research. 
As a definition of empathy between an anthropomorphic agent or robot and a human, Paiva represented empathy agents in two different ways and illustrated them \cite{Paiva04,Paiva11,Paiva17}: A) targets to be empathized with by humans and B) observers who empathize with humans. 
In this study, we use the empathic target agent to promote human empathy.

\section{Experimental methods}
\subsection{Experimental goals and design}
The purpose of this study is to investigate whether human empathy toward an agent is affected by the agent's reaction and human's preference during interaction with an robotic virtual agent (RVA).
It will then investigate whether agents can be forgiven when they make mistakes.
We believe that this research will facilitate the use of agents in human society by influencing human empathy.
In addition, knowing the factors that allow agents to make mistakes will be useful for future use of agents in society.
For these purposes, we formulated two hypotheses.

\begin{enumerate}
\item[\textbf{H1}:] Agent's reaction and human's preference affect human empathy toward agents.
\item[\textbf{H2}:] Agent's reaction and human's preference influence when humans accept agents' mistakes.
\end{enumerate}

We arrived at this hypothesis because previous studies have shown that agent reactions (facial expressions and gestures) affect human empathy.
In addition, the intention of the agent's human's preference was to investigate whether the results of human selections affect empathy toward an agent by adding an element that allows humans to interact with the agent.

Similarly, this study focuses on whether agents' mistakes are acceptable. 
This study investigates an agent's relationship with a human in situations where the agent is wrong. 
If the agent's reaction and human's preference affect how a human accepts the agent's mistakes, then there is no need to incorporate empathy toward the agent in order to maintain the human's impression of the agent.

To test these hypotheses, an experiment was conducted with a three-factor mixed design with three factors: agent reaction, human's preference, and pre- and post-task. 
The levels between participants were 2 (available, not available) for agent reaction and 2 (available, not available) for human's preference. 
The within-participants level was 2 pre- and post-task. 
Participants participated in only one of the four different content conditions. 
The dependent variable was the questionnaire that participants responded to (empathy, tolerance for error, other).

\subsection{Experimental details}
The experiment was conducted in an online environment. 
The environment used is already a common method of experimentation\cite{Davis99,Crump13,Okamura20}. 
As mentioned earlier, the goal of this study is to promote human empathy toward RVA. 
A scheduling agent was also used in this study to measure the acceptance of the agent's mistakes.
For this reason, we believed that the same effect as being face to face could be achieved even in an online environment.

Before performing the task, a questionnaire was administered to measure empathy toward RVA. 
At the same time, another questionnaire was administered to determine whether participants could accept the agent's mistakes. 
At this time, participants were not allowed to see the agent's reactions or to select the color of the agent. 
This questionnaire was administered before the task in order to see the effect of participants' empathy toward the agent and the change in the acceptance of the agent's mistakes.

\begin{figure}[tbp]
		\begin{center}
		\includegraphics[scale=0.3]{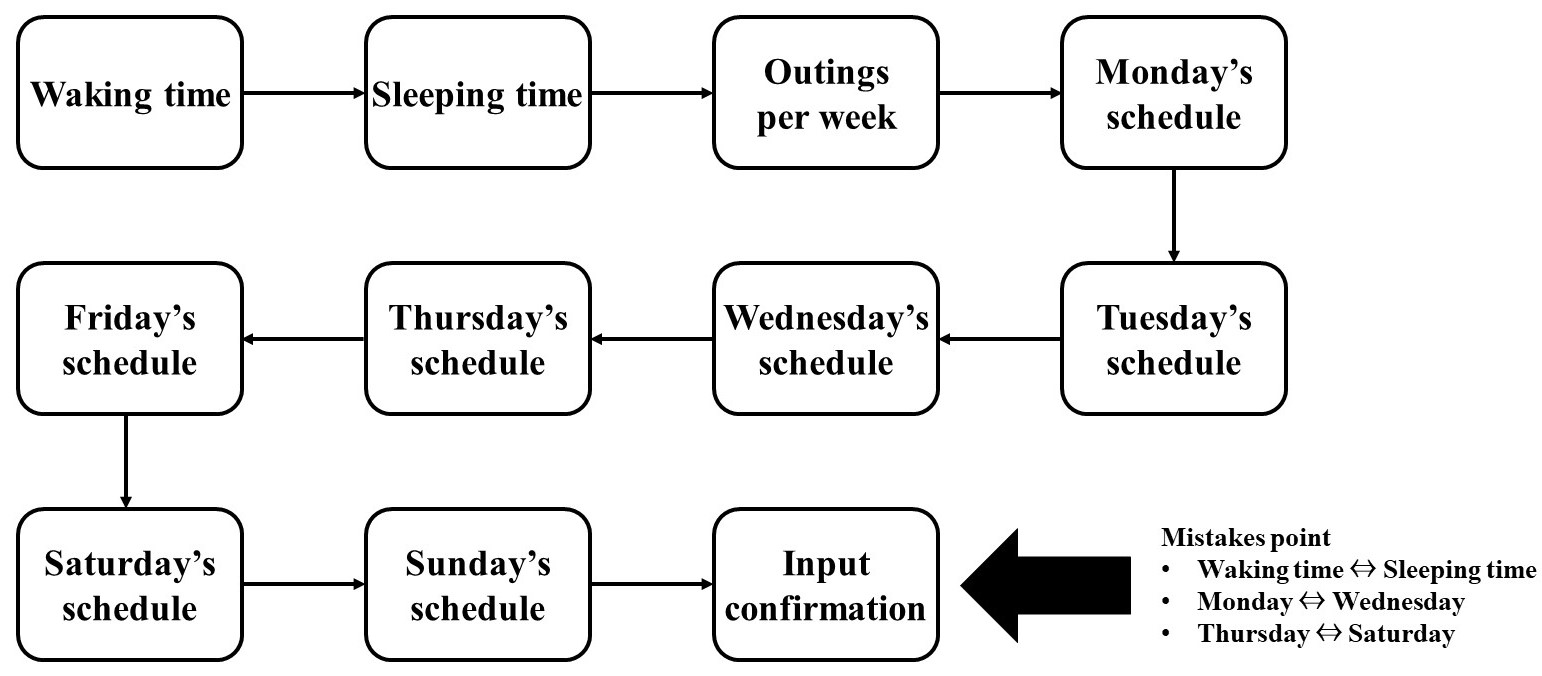}
		\caption{Flowchart of the schedule}
		\label{flowchart}
	\end{center}
\end{figure}

Participants selected an agent by color from among multiple differently colored agents before beginning the scheduling task. 
In the no human's preference condition, participants were told that the agent displayed would manage a schedule. 
The schedule consisted of 10 items: the participant's weekly schedule, waking time, sleeping time, and number of outings per week. 
Fig.~\ref{flowchart} is a flowchart of the schedule entry order and up to the confirmation screen.
During the scheduling task, the agent exhibited several reactions to the input information. 
The scheduling agent was designed to remember the participant's schedule but to make sure that the agent made a mistake when participant checked the schedule for the last time. 
There were three areas where mistakes were made: 
(1) the waking and sleeping times were reversed, the (2) Monday and Wednesday schedules were reversed, and (3) the schedules for Thursday and Saturday were reversed.
Waking and sleeping times and the number of outings per week were selected from a list of options, and schedules from Monday to Sunday were answered in the form of free-text responses.

This was done to investigate how an agent's mistakes affect human empathy and acceptance. 
To screen out unfair participants, participants reported whether the schedule was correct or incorrect at the last confirmation of the schedule. 
Only those participants who reported that the schedule was wrong were subsequently administered the same questionnaire about RVA as before the task. 
Two additional questions were also asked. 
Finally, they were asked to write their impressions of the experiment in free text.

\subsection{Experimental environment and Participants}
Participants were recruited for the experiment using a Yahoo! crowdsourcing company.
They were paid 55 yen after completing all tasks as a reward for participating.
A website was created for the experiment, which was limited to using a PC.

There were a total of 197 participants. 
The average age was 48.82 years (standard deviation: 11.08), with a minimum of 19 years and a maximum of 77 years.
The gender breakdown was 144 males and 53 females. 
The number of participants in each condition is shown in Table~\ref{table1}.
We then applied Cronbach's $\alpha$ coefficient to determine the reliability of the questionnaire responses, which was found to be between 0.4040 and 0.8190 in all conditions. 
Some participant groups had lower Cronbach's $\alpha$ values, but we also found several conditions with values between 0.7 and 0.8. Therefore, we used the questionnaire without any modifications.

\begin{table}[tbp]
\caption{Number of participants in each condition}
\centering
\begin{tabular}{cc|cc} \hline
 & & \multicolumn{2}{l}{Reaction} \\ \cline{3-4} 
 & & Yes & No \\ \hline
 \multicolumn{1}{l|}{} & Yes & 50 & 50 \\
 \multicolumn{1}{l|}{human's preference} & No & 49 & 48 \\ \hline
\end{tabular}
\label{table1}
\end{table}

\subsection{Agent's reactions}
\begin{figure}[tbp]
		\begin{center}
		\includegraphics[scale=0.18]{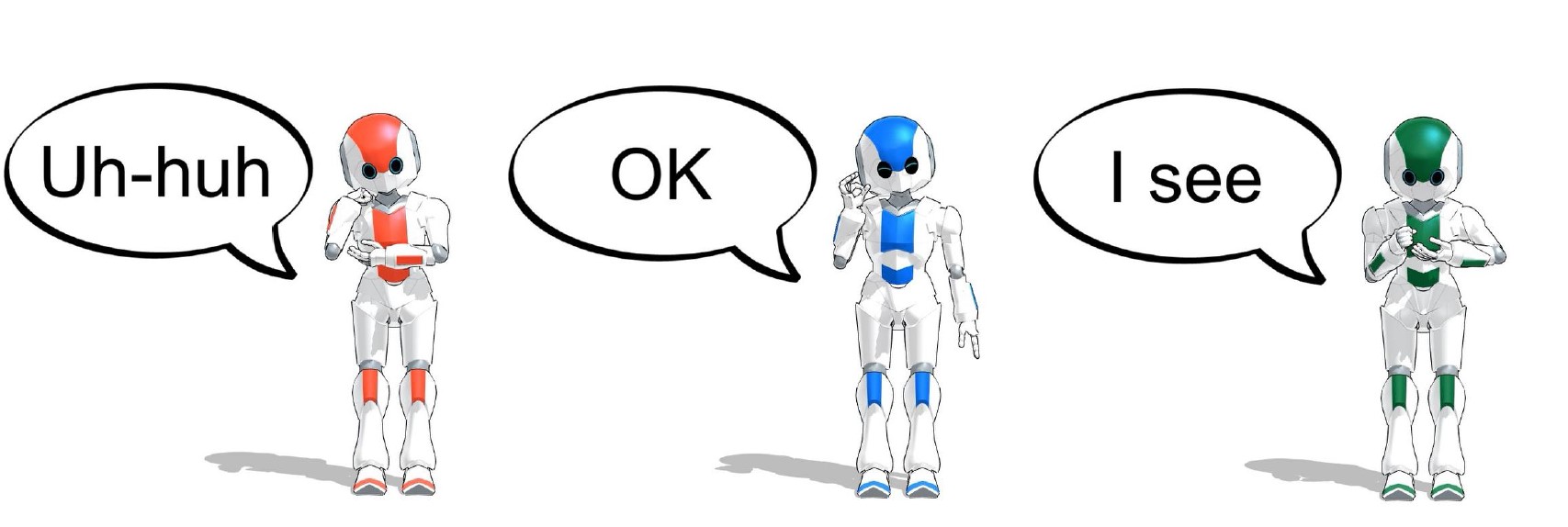}
		\caption{Types of agent reactions}
		\label{reaction}
	\end{center}
\end{figure}

In this study, two levels of agent reactions were prepared.
In the one with the agent's reaction, RVA responded with gestures and comments when the participant's schedule was sent. 
For the one without the agent's reaction, it did not respond to the participant's schedule and stayed upright. 
The three types of reactions actually seen by the participants are shown in Fig.~\ref{reaction}. 
The three types were displayed three times equally and in the order in which they were displayed to the participants. 
Because the reactions were based on the content of the participant's schedule, RVA did not react to anything when the schedule was first entered.

There is a study by Tsumura and Yamada~\cite{Tsumura23-2} as an example of how agent representations affected human empathy toward agents, but unlike reactions, which are very brief representations, they did not focus on detailed representations for a single action, as in this experiment.

\subsection{Human's preference} 
\begin{figure}[tbp]
		\begin{center}
		\includegraphics[width=86mm]{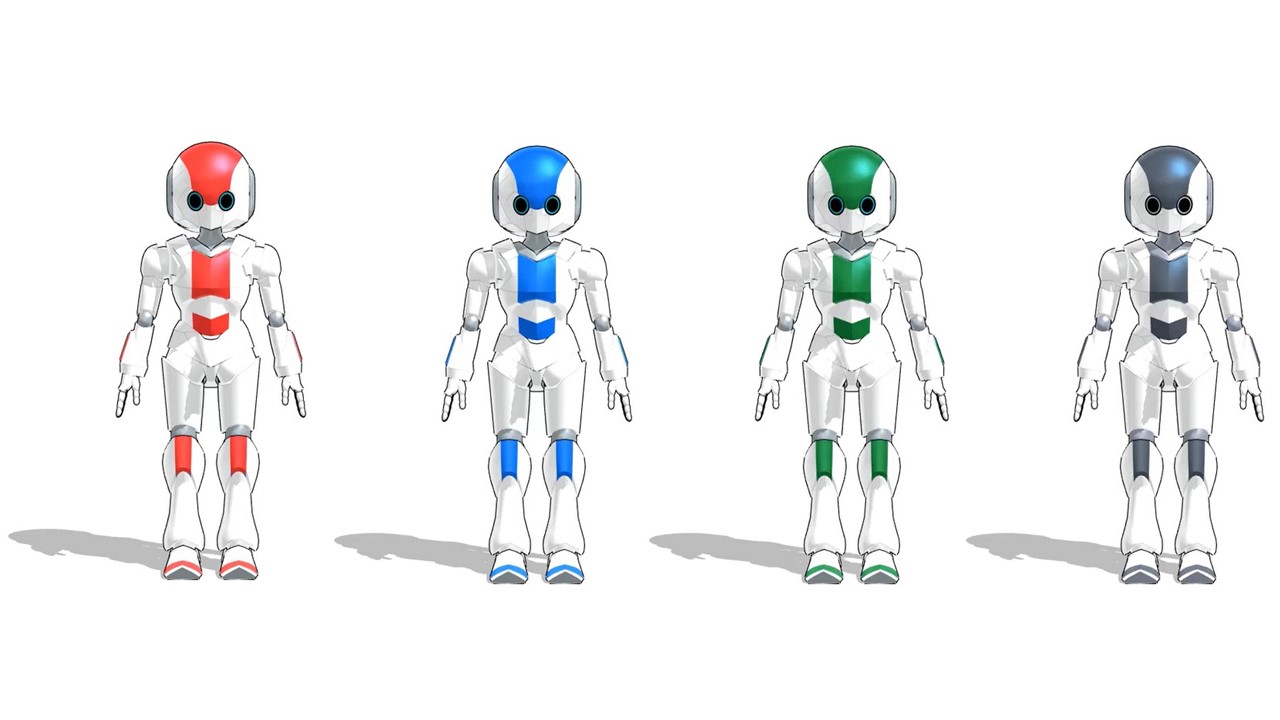}
		\caption{Types of agent's color}
		\label{color}
	\end{center}
\end{figure}

There were two levels for a human's preference factor. 
In this study, participants were asked to choose the color of the agent's appearance to investigate people's preferences.
If a human's preference was had, the participants were required to choose one of three types of RVA: red, blue, or green.
There was no difference in the agent's personality or behavior based on human's preference.
If no human's preference was had, the gray agent would manage the schedule. 
The color of each agent is shown in Fig.~\ref{color}. 

In this study, we did not consider bias in human's preference. 
The human's preference factor was chosen as a factor that allowed participants to interact with RVA. 
The purpose was to investigate whether the participants' impressions of RVA changed depending on the selection.
Therefore, differences in participants' feelings towards particular colors and gender bias were ignored in this study.

\subsection{Questionnaire}
Participants completed a questionnaire before and after the task. 
In this study, human empathy for the agent was evaluated based on changes in human empathy characteristics.
The questionnaire was a 12-item questionnaire modified from the Interpersonal Reactivity Index (IRI), which is used to investigate the characteristics of empathy, to suit the present experiment \cite{Davis80}. 
The modified questionnaire has already been used in several previous studies by Tsumura and Yamada~\cite{Tsumura22,Tsumura23-1,Tsumura23-2}.
The two questionnaires before and after were the same. Both were based on the IRI and were surveyed on a 5-point Likert scale (1: not applicable, 5: applicable). 
The questionnaire used is shown in Table~\ref{table2}. Q4, Q9, and Q10 are inverted items, so the scores were reversed when analyzing them.

Three questions other than those related to empathy were prepared: QA, QB, and QC. QA was surveyed before and after the task, while QB and QC were surveyed only after the task.
QA was investigated before and after the task to compare the difference between the participants' assumed tolerance of agent error and the actual agent error after it occurred.
QB was prepared to investigate whether the agent's very brief reactions appeared emotional.
QC was an item to investigate the impact of differences in factors on the participants' evaluations of the agent's future use.
These three questions were also surveyed on a 5-point Likert scale (1: not applicable, 5: applicable).
The questionnaire is shown in Table~\ref{table2}.

\begin{table*}[tbp] 
    \caption{Summary of questionnaire used in this experiment}
    \centering
    \scalebox{0.8}{
    \begin{tabular}{l}\hline 
        \textbf{Affective empathy}\\ \hline
        \textbf{Personal distress}\\
        Q1: If an emergency happens to the character, you would be anxious and restless.\\
        Q2: If the character is emotionally disturbed, you would not know what to do.\\
        Q3: If you see the character in need of immediate help, you would be confused and would not know what to do.\\
        \textbf{Empathic concern}\\
        Q4: If you see the character in trouble, you would not feel sorry for that character.\\
        Q5: If you see the character being taken advantage of by others, you would feel like you want to protect that character.\\
        Q6: The character's story and the events that have taken place move you strongly.\\\hline
        \textbf{Cognitive empathy}\\ \hline
        \textbf{Perspective taking}\\
        Q7: You look at both the character's position and the human position.\\
        Q8: If you were trying to get to know the character better, you would imagine how that character sees things.\\
        Q9: When you think you're right, you don't listen to what the character has to say.\\
        \textbf{Fantasy scale}\\
        Q10: You are objective without being drawn into the character's story or the events taken place.\\
        Q11: You imagine how you would feel if the events that happened to the character happened to you.\\
        Q12: You get deep into the feelings of the character.\\\hline
        \textbf{Other questions than empathy}\\ \hline
        QA: If the scheduling agent makes a mistake, can you forgive the mistake?\\ 
        QB: Did the scheduling agent express emotions?\\ 
        QC: Would you like to use a scheduling agent in the future?\\ 
        \hline
    \end{tabular}}
    \label{table2}
\end{table*}

\subsection{Analysis method}
The analysis was a three-factor analysis of variance (ANOVA). 
The between-participant factors were the two levels of agent's reaction and two levels of agent's human's preference. 
The within-participants factor consisted of two levels of empathy values before and after the task.
On the basis of the results of the participants' questionnaires, we investigated how the agent's reaction and human's preference influenced the promotion of empathy as factors that elicit human empathy. 
The numerical values of empathy aggregated before and after the task were used as the dependent variable. 
Three of our own questions were also used as dependent variables. R (R ver. 4.1.0) was used for the ANOVA. 

\section{Results}
\begin{table*}[tbp]
 \caption{Results of participants' statistical information}
 \centering
 \scalebox{0.9}{
 \begin{tabular}{c|c|c|cc||c|c|c|cc}\hline 
 \multicolumn{2}{c|}{Category} & Conditions & Mean & S.D. & \multicolumn{2}{c|}{Category} & Conditions & Mean & S.D. \\ \hline 
 & & reaction-preference & 36.94 & 5.705 & & & reaction-preference & 2.660 & 0.8947 \\ 
 & & reaction-no preference & 38.78 & 5.363 & & & reaction-no preference & 2.837 & 0.7457\\ 
 Empathy & pre & no reaction-preference & 36.90 & 6.370 & Agent's & pre & no reaction-preference & 2.880 & 0.8722 \\ 
 & & no reaction-no preference & 38.67 & 6.096 & acceptance & & no reaction-no preference & 2.563 & 0.9655 \\ \cline{2-5}\cline{7-10}
 & & reaction-preference & 33.98 & 7.347 & & & reaction-preference & 2.560 & 0.9510 \\ 
 (Q1-Q12) & & reaction-no preference & 34.94 & 6.710 & (QA) & & reaction-no preference & 2.755 & 0.7781 \\ 
 & post & no reaction-preference & 34.72 & 6.716 & & post & no reaction-preference & 2.600 & 1.107 \\ 
 & & no reaction-no preference & 35.71 & 6.694 & & & no reaction-no preference & 2.271 & 0.9618 \\ \hline
 & & reaction-preference & 2.840 & 0.9116 & & & reaction-preference & 2.140 & 0.9260 \\ 
 Expressed & & reaction-no preference & 2.674 & 1.068 & Continued & & reaction-no preference & 2.286 & 0.9129 \\ 
 emotions & post & no reaction-preference & 2.000 & 0.8571 & use & post & no reaction-preference & 2.080 & 0.9223 \\ 
 (QB) & & no reaction-no preference & 1.896 & 0.7217 & (QC) & & no reaction-no preference & 1.979 & 0.8627 \\ \hline
 \end{tabular}} \\ 
 \label{table3}
\end{table*}

\begin{table*}[tbp]
\caption{Analysis results of ANOVA}
\centering
\scalebox{1.0}{
\begin{tabular}{c|llll}\hline
& \multicolumn{1}{c}{Factor} & \multicolumn{1}{c}{\em{F}} & \multicolumn{1}{c}{\em{p}} & \multicolumn{1}{c}{$\eta^2_p$}\\ \hline
& Reaction & 0.1567 & 0.6926 \em{ns} & 0.0008 \\ 
& human's preference & 2.606 & 0.1081 \em{ns} & 0.0133\\
Empathy & Before/after task & 94.11 & 0.0000 *** & 0.3278 \\ 
(Q1-12)& Reaction $\times$ human's preference & 0.0001 & 0.9909 \em{ns} & 0.0000 \\ 
& Reaction $\times$ Before/after task & 1.817 & 0.1793 \em{ns} & 0.0093 \\ 
& human's preference $\times$ Before/after task & 1.810 & 0.1801 \em{ns} & 0.0093 \\
& Reaction $\times$ human's preference $\times$ Before/after task & 0.0064 & 0.9363 \em{ns} & 0.0000 \\ 
\hline
& Reaction & 1.132 & 0.2887 \em{ns} & 0.0058 \\ 
& human's preference & 0.3440 & 0.5582 \em{ns} & 0.0018\\
Agent's & Before/after task & 10.68 & 0.0013 ** & 0.0525 \\ 
acceptance & Reaction $\times$ human's preference & 4.725 & 0.0309 * & 0.0239 \\ 
(QA)& Reaction $\times$ Before/after task & 2.864 & 0.0922 \em{ns} & 0.0146 \\ 
& human's preference $\times$ Before/after task & 0.0008 & 0.9768 \em{ns} & 0.0000 \\
& Reaction $\times$ human's preference $\times$ Before/after task & 0.0170 & 0.8964 \em{ns} & 0.0001 \\ 
\hline
Expressed & Reaction & 39.86 & 0.0000 *** & 0.1712 \\ 
emotions & human's preference & 1.116 & 0.2921 \em{ns} & 0.0057\\ 
(QB) & Reaction $\times$ human's preference & 0.0592 & 0.8080 \em{ns} & 0.0003 \\ 
\hline
Continued & Reaction & 2.012 & 0.1577 \em{ns} & 0.0103 \\ 
use & human's preference & 0.0302 & 0.8623 \em{ns} & 0.0002\\ 
(QC) & Reaction $\times$ human's preference & 0.9100 & 0.3413 \em{ns} & 0.0047 \\ 
\hline
\end{tabular}} \\
\hspace{-75mm}
\em{p}:
{{*}p\textless\em{0.05}}
{{**}p\textless\em{0.01}}
{{***}p\textless\em{0.001}}
\label{table4}
\end{table*}

\begin{table*}[tbp]
\caption{Analysis results of simple main effect}
\centering
\scalebox{1.0}{
\begin{tabular}{c|llll}\hline
& \multicolumn{1}{c}{Factor} & \multicolumn{1}{c}{\em{F}} & \multicolumn{1}{c}{\em{p}} & \multicolumn{1}{c}{$\eta^2_p$}\\ \hline
& Reaction with preference & 0.5613 & 0.4555 \em{ns} & 0.0057 \\ 
Agent's & Reaction with no preference & 5.851 & 0.0175 * & 0.0580\\ 
acceptance & human's preference with reaction & 1.577 & 0.2122 \em{ns} & 0.0160 \\
(QA) & human's preference with no reaction & 3.160 & 0.0786 + & 0.0319 \\ 
\hline
\end{tabular}} \\
\hspace{-50mm}
\em{p}:
{{+}p\textless\em{0.1}}
{{*}p\textless\em{0.05}}
{{**}p\textless\em{0.01}}
\label{table5}
\end{table*}

\begin{figure*}[tbp]
		\begin{center}
		\includegraphics[scale=0.35]{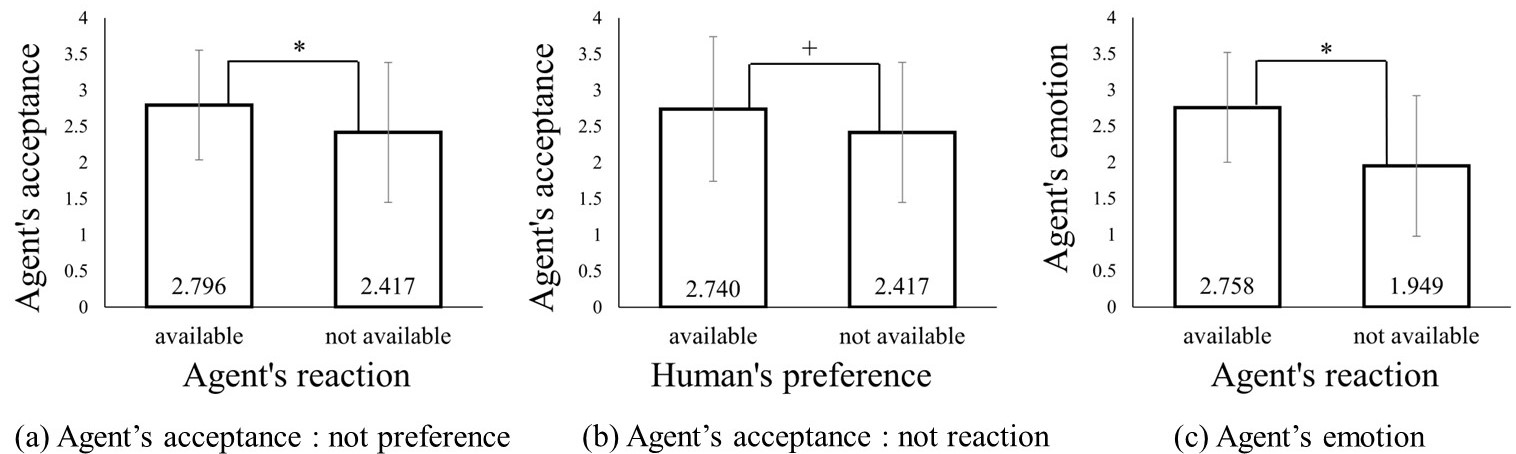}
		\caption{Result of each main effect or simple main effect}
		\label{exp1}
	\end{center}
\end{figure*}

All 12 questionnaire items were analyzed together. 
Table~\ref{table3} shows the statistical results for each. 
Table~\ref{table4} shows the results of each ANOVA. 

To begin, as can be seen from Table~\ref{table4}, when the participants' empathy for the agent was examined, there were no significant differences other than the main pre- and post-task effects.
Comparing the pre- and post-task empathy values in Table~\ref{table3} for the ability to empathize with the agent, empathy decreased after the task (all pre-task: mean = 37.81, S.D. = 5.921; all post-task: mean = 34.83, S.D. = 6.850).

The results for agent tolerance showed an interaction between the agent's reaction factor and human's preference factor. 
Therefore, we analyzed the simple main effect in Table~\ref{table5} and found that the agent's reaction more likely lead to tolerance to the agent's error when the participant did not make a human's preference. 
We also found that, although a significant trend, the human's preference made its mistake more acceptable when there was no agent reaction.
These results are shown in Fig.~\ref{exp1}(a) and (b).

The results for the agent's emotional expression showed a main effect for the agent's reaction factor. 
However, no main effect was found for the agent's human's preference factor. 
This indicated that the agent's reactions appeared to be emotionally charged.
These results are shown in Fig.~\ref{exp1}(c).
Finally, there were no significant differences in the continued use of the agents in all conditions.

\section{Discussion}
\subsection{Supporting hypotheses}
This experiment was designed to investigate the conditions necessary for humans to empathize with anthropomorphic agents.
In particular, by investigating when an agent makes a mistake on a task, the goal was to identify factors that influence empathy between an agent who makes a mistake on a task and a human.
To this end, two hypotheses were formulated, and the data obtained from the experiment were analyzed.

The results supported one hypothesis, but not the other. 
In \textbf{H1}, we thought that the agent's reaction and human's preference would affect the participants' empathy toward the agent, but this one was not supported.
In the present experiment, there was a decrease in empathy after the task in all conditions. 
This was also the case in Tsumura and Yamada~\cite{Tsumura22}.
The reason for the decrease in empathy may be that the all agents made mistakes, as there was no significant difference in each factor.
Therefore, as a future study, we will compare the results with those obtained when the agents did not make mistakes.
\\ \indent
In \textbf{H2}, we thought that the agent's reaction and human's preference would affect the tolerance toward the mistakes the agent made, which was supported here.
In each case, when the agent's reaction was absent, the agent's mistake was accepted when human's preference was present, and when it was absent, the agent's mistake was accepted when the agent's reaction was present. 
On the other hand, when both agent reaction and human's preference were present, there was no effect. 
A possible reason for the lack of acceptance of agent error when both conditions were included is discouragement toward the agent. 
Therefore, as a future study, we will compare the results with those obtained when the agent does not make mistakes.
\\ \indent
We also investigated the agent's emotional expressions and found that the agent's reactions appeared to be emotional expressions. 
However, regardless of the agent's reaction, empathy toward the agent was reduced, indicating that even when the agent acts emotionally, it is unlikely to affect empathy in situations where the agent makes a mistake on the task.

\subsection{Limitations}
One limitation of this experiment is that by eliminating factors other than the agent's reaction and the human's preference, the task itself was perceived as tedious, and the simplicity of the task may have reduced empathy.
By not allowing RVA to engage in conversation or introduce themselves beyond the scheduling task, participants may have decreased their impression of RVA.
Also, RVA were silent in this experiment, which was also done to eliminate the effect of voice on empathy and thus simplified the agents' reactions.

Although there was no need for an in-person experiment in this experiment, an in-person experiment using actual equipment could have made a difference in the impact on participants' impressions.
A scheduling task was used in this study to investigate whether humans empathize with RVA and accept their mistakes even when they make mistakes on a task. 
However, even when agents' mistakes are acceptable, it is necessary to investigate the extent to which they are acceptable.

\section{Conclusions}
In this study, we investigated agent reaction and human's preference, focusing on human-agent task error as a factor that causes humans to empathize with RVA.
RVA was designed to be in charge of managing the human's schedule and to make some mistakes in the input information.
Two hypotheses were formulated and tested.
The results showed that empathy toward RVA decreased when RVA made a mistake on the task.
In addition, agent reaction and human's preference were shown to be effective in helping humans accept agent mistakes. 
However, it was shown that the use of either factor was not effective. 
Future research should investigate empathy and acceptance toward agents when they do not make mistakes on a task since it was confirmed that empathy toward RVA decreases when they make mistakes on a task.

\section*{Acknowledgments}
This work was partially supported by JST, CREST (JPMJCR21D4), Japan.
This work was also supported by JST, the establishment of university fellowships towards the creation of science technology innovation, Grant Number JPMJFS2136.

%
%
%
\bibliographystyle{splncs04}
\bibliography{test}

\end{document}